\documentclass[prl,showpacs,superscriptaddress,twocolumn,floatfix]{revtex4}
\usepackage{graphicx,dcolumn}
\begin{document}
\title{\boldmath Recoil polarization and beam-recoil double
       polarization measurement of $\eta$ electroproduction on the
       proton in the region of the $S_{11}(1535)$ resonance}
\author{H.~Merkel}
\email{Merkel@kph.uni-mainz.de}
\homepage[\\URL: ]{http://wwwa1.kph.uni-mainz.de/}
\affiliation{Institut f\"ur Kernphysik,
  Johannes Gutenberg-Universit\"at Mainz, D-55099~Mainz, Germany}
\author{P.~Achenbach}\affiliation{Institut f\"ur Kernphysik,
  Johannes Gutenberg-Universit\"at Mainz, D-55099~Mainz, Germany}
\author{C.~Ayerbe~Gayoso}\affiliation{Institut f\"ur Kernphysik,
  Johannes Gutenberg-Universit\"at Mainz, D-55099~Mainz, Germany}
\author{J.\,C.~Bernauer}\affiliation{Institut f\"ur Kernphysik,
  Johannes Gutenberg-Universit\"at Mainz, D-55099~Mainz, Germany}
\author{R.~B\"ohm}\affiliation{Institut f\"ur Kernphysik,
  Johannes Gutenberg-Universit\"at Mainz, D-55099~Mainz, Germany}
\author{D.~Bosnar}\affiliation{Department of Physics, University of Zagreb,
  HR-10002 Zagreb, Croatia}       
\author{B.~Cheymol}\thanks{Supported by the French CNRS/IN2P3}
\affiliation{Laboratoire de Physique Corpusculaire IN2P3-CNRS,
  Universit\'e Blaise Pascal, F-63170 Aubi\`ere Cedex, France}
\author{M.\,O.~Distler}\affiliation{Institut f\"ur Kernphysik,
  Johannes Gutenberg-Universit\"at Mainz, D-55099~Mainz, Germany}
\author{L.~Doria}\affiliation{Institut f\"ur Kernphysik,
  Johannes Gutenberg-Universit\"at Mainz, D-55099~Mainz, Germany}
\author{H.~Fonvieille}\thanks{Supported by the French CNRS/IN2P3}
\affiliation{Laboratoire de Physique Corpusculaire IN2P3-CNRS, 
  Universit\'e Blaise Pascal, F-63170 Aubi\`ere Cedex, France}
\author{J.~Friedrich}\affiliation{Institut f\"ur Kernphysik,
  Johannes Gutenberg-Universit\"at Mainz, D-55099~Mainz, Germany}
\author{P.~Janssens}
\thanks{Ph. D. fellowship Research Foundation - Flanders (FWO)}
\affiliation{Department of Subatomic and Radiation Physics,
  University of Ghent, B-9000 Ghent, Belgium}
\author{M.~Makek}\affiliation{Department of Physics, University of Zagreb,
  HR-10002 Zagreb, Croatia}
\author{U.~M\"uller}\affiliation{Institut f\"ur Kernphysik,
  Johannes Gutenberg-Universit\"at Mainz, D-55099~Mainz, Germany} 
\author{L.~Nungesser}\affiliation{Institut f\"ur Kernphysik,
  Johannes Gutenberg-Universit\"at Mainz, D-55099~Mainz, Germany}
\author{J.~Pochodzalla}\affiliation{Institut f\"ur Kernphysik,
  Johannes Gutenberg-Universit\"at Mainz, D-55099~Mainz, Germany} 
\author{M.~Potokar}
\affiliation{Jo\v{z}ef Stefan Institute, SI-1001 Ljubljana, Slovenia}
\author{S.~S\'anchez~Majos}\affiliation{Institut f\"ur Kernphysik,
  Johannes Gutenberg-Universit\"at Mainz, D-55099~Mainz, Germany} 
\author{B.\,S.~Schlimme}\affiliation{Institut f\"ur Kernphysik,
  Johannes Gutenberg-Universit\"at Mainz, D-55099~Mainz, Germany}
\author{S.~\v{S}irca}
\affiliation{Jo\v{z}ef Stefan Institute, SI-1001 Ljubljana, Slovenia}
\affiliation{Department of Physics, University of Ljubljana, 
  SI-1000 Ljubljana, Slovenia} 
\author{L.~Tiator}\affiliation{Institut f\"ur Kernphysik,
  Johannes Gutenberg-Universit\"at Mainz, D-55099~Mainz, Germany}
\author{Th.~Walcher}\affiliation{Institut f\"ur Kernphysik,
  Johannes Gutenberg-Universit\"at Mainz, D-55099~Mainz, Germany} 
\author{M.~Weinriefer}\affiliation{Institut f\"ur Kernphysik,
  Johannes Gutenberg-Universit\"at Mainz, D-55099~Mainz, Germany}
\collaboration{A1 Collaboration}
\noaffiliation
\date{24 May, 2007}
\begin{abstract}
  The beam-recoil double polarization $P_{x'}^h$ and $P_{z'}^h$ and
  the recoil polarization $P_{y'}$ were measured for the first time
  for the $p(\vec{e},e'\vec{p}\,)\eta$ reaction at a four-momentum
  transfer of $Q^2=0.1\,\mathrm{GeV}^2/c^2$ and a center of mass
  production angle of $\theta = 120^\circ$ at MAMI C. With a center of
  mass energy range of $1500\,\mathrm{MeV} < W < 1550\,\mathrm{MeV}$
  the region of the $S_{11}(1535)$ and $D_{13}(1520)$ resonance was
  covered. The results are discussed in the framework of a
  phenomenological isobar model (Eta-MAID). While $P_{x'}^h$ and
  $P_{z'}^h$ are in good agreement with the model, $P_{y'}$ shows a
  significant deviation, consistent with existing photoproduction data
  on the polarized-target asymmetry.
\end{abstract}
\maketitle


The electromagnetic production of $\eta$ mesons is a selective probe
to study the resonance structure of the nucleon. Since the $\eta$
meson has isospin $I=0$ only nucleon resonances with isospin $I=1/2$
contribute to the reaction, opening a unique window to small
resonances which are buried in the case of pion production and $\pi N$
scattering by large $I=3/2$ resonances. In addition, due to the small
$\eta NN$ coupling, the non-resonant background is strongly suppressed
and the resonance excitation can be studied in a clean way.


A vast amount of unpolarized photoproduction data
\nocite{Prepost:1967} \nocite{Bacci:1968} \nocite{Bloom:1968}
\nocite{Delcourt:1969dn} \nocite{Krusche:1995nv}
\nocite{Dytman:1995vm} \nocite{Renard:2002} \nocite{Dugger:2002ft}
\nocite{Crede:2005} \nocite{Nakabayashi:2006ut} \cite{Prepost:1967,
Bacci:1968, Bloom:1968, Delcourt:1969dn, Krusche:1995nv,
Dytman:1995vm, Renard:2002, Dugger:2002ft, Crede:2005,
Nakabayashi:2006ut}, the more recent ones with impressive accuracy,
established the dominance of the $s$-wave in the threshold region.
Most authors interpret this fact with a reaction mechanism dominated
by the $S_{11}(1535)$ resonance. Phenomenological isobar models
\cite{Knochlein:1995qz, Chiang:2001as, Aznauryan:2004jd,
Anisovich:2005tf} can successfully describe the data with a standard
Breit-Wigner shape of the contributing resonances, although dividing 
out the phase space reveals a relatively flat energy dependence of
the $s$-wave amplitude at threshold.

To further disentangle resonances with small couplings to the $\eta N$
channel beyond the pure $s$-wave production, polarization observables
are indispensable. The polarized target asymmetry was measured in Bonn
at the PHOENICS experiment \cite{Bock:1998rk}. This measurement showed
a surprising angular structure, which cannot be described by the
existing phenomenological models. A detailed model-independent study
\cite{Tiator:1999gr} showed, that one possibility to describe these
data is to include a strong phase shift between $s$- and $d$-waves,
basically giving up the standard Breit-Wigner phase for either the
$S_{11}(1535)$ or for the $D_{13}(1520)$ resonance. The somewhat
arbitrary introduction of such a phase shift was chosen to ensure
that the differential cross section data are still well described by
the model. However, since the error bars in ref.\ \cite{Bock:1998rk}
are quite large, this discrepancy was disputed for a long time and is
still an open issue.

Other polarization measurements were not sensitive to the same
multipole interference. A first measurement of the recoil polarization
in 1970 \cite{Heusch:1970tr} covered only a center of mass angle of 90
degrees, where this interference is zero. At GRAAL the photon beam
asymmetry has been measured \cite{Ajaka98} and contributions from the
$D_{13}(1520)$ and $F_{15}(1680)$ were established. Recent
measurements of the polarized-beam asymmetry at ELSA
\cite{Elsner:2007} are sensitive to the real part, but not to the
imaginary part of the interference amplitude. A pioneering experiment
on the photon-target double polarization asymmetry at MAMI
\cite{Ahrens:2003bp} could only confirm the $s$-wave dominance.

On the theory side calculations of $\eta$ photoproduction have been
performed in isobar models \cite{Knochlein:1995qz, Chiang:2001as,
Aznauryan:2004jd, Anisovich:2005tf}, in the quark model
\cite{Saghai:2001yd}, with dispersion relations
\cite{Aznauryan:2003zg}, with coupled channels \cite{Feuster:1998cj},
and also a partial-wave analysis \cite{SAID} has been performed. With
the increasing database of high-quality data the analyses became more
reliable and most of the data are very well described. However, none
of these model calculations and partial-wave analyses were able to
explain the target polarization asymmetry seen in the Bonn
experiment. In fact, the phase shift found in the model-independent
approach \cite{Tiator:1999gr} goes beyond the usual approaches, where
resonances are treated as nucleon isobars. An alternative way to look
for such an unusual phase would be the dynamical approaches. So far
such unitary approaches \cite{Kaiser:1996js, Inoue:2001ip} could
successfully describe the $S_{11}$ partial wave even without a
nucleon isobar by chiral dynamics with coupled channels. However, the
interference with other channels such as $D_{13}$ has not yet been
studied in this framework.

The aim of this work was to test the possibility of the phase shift
in an independent experiment. By choosing recoil polarization and
beam-recoil double polarization observables we were sensitive to the
same interference of multipoles as tested by the polarized-target
asymmetry, as will be shown in the next section. This experiment was
performed with a four-momentum transfer of $Q^2 =
0.1\,\mathrm{GeV^2}/c^2$. Previous electroproduction experiments
\cite{Armstrong:1998wg, Thompson:2000by, Denizli:2007tq} showed
already, that the $Q^2$ dependence of the cross section is flat and
the contribution of longitudinal multipoles is small, so that the
phenomenological models are considered reliable for the extrapolation
from the photon point to this small $Q^2$ value.


The cross section for polarized electroproduction of pseudoscalar
mesons can be written in terms of structure functions as (see ref.\
\cite{Knochlein:1995qz} for full notation)
\newlength\strl\settoheight\strl{$\sqrt{2 \epsilon ( 1 + \epsilon)}$}
\begin{eqnarray*}
\label{dsigmafull}
\frac{d^5\sigma}{dE'd\Omega'd\Omega} &=& \Gamma \frac{d\sigma}{d\Omega},\\[2mm]
\frac{d\sigma}{d\Omega} & = & \frac{|\bf q|}{k_w}P_{\alpha}P_{\beta}
\left\{\rule{0mm}{\strl}R_{\rm T}^{\beta\alpha}+\epsilon R_{\rm L}^{\beta\alpha}\right.\\
&&+\sqrt{2\epsilon(1+\epsilon)}
  (^c\!R_{\rm TL}^{\beta\alpha}\cos\phi+^s\!\!R_{\rm TL}^{\beta\alpha}\sin\phi)\\
&&+\epsilon(^c\!R_{\rm TT}^{\beta\alpha}
  \cos2\phi+^s\!\!R_{\rm TT}^{\beta\alpha}\sin2\phi)\\
&&+h\sqrt{2\epsilon(1-\epsilon)}
  (^c\!R_{\rm TL'}^{\beta\alpha}\cos\phi+^s\!\!R_{\rm TL'}^{\beta\alpha}\sin\phi)\\
&&\left.+h\sqrt{1-\epsilon^2}R_{\rm TT'}^{\beta\alpha}\right\},
\end{eqnarray*}
where $h$ is the longitudinal beam polarization, the index $\alpha =
0,x,y,z$ denotes the direction of the target polarization $P_\alpha$, and
$\beta=0,x',y',z'$ denotes the direction of the recoil polarization
$P_\beta$ in the center of mass frame with $z'$ in direction of the
$\eta$, $y'$ perpendicular to the p-$\eta$-plane and $x'\times
y'=z'$. $\Gamma$ is the usual virtual photon flux, $\epsilon$ the
transverse polarization of the virtual photon and $\theta$ and $\phi$
the center of mass angles of the outgoing $\eta$ with respect to the
photon direction. $k_w$ is the equivalent real photon energy in the
c.m. frame and $\bf q$ is the $\eta$ c.m. momentum.

With polarized beam, unpolarized target, and an in-plane
(\textit{i.e.} $\phi=0,\pi$) measurement of the recoil polarization,
one can measure in addition to the unpolarized cross section two
helicity-dependent polarizations and one helicity-independent
polarization
\begin{eqnarray*}
\sigma_0         & = & \frac{|{\bf q}|}{k_w}
                       \left\{\rule{0mm}{\strl}
                       R_{\rm T}^{00} + \epsilon R_{\rm L}^{00}\right.\\
                 &   & \left.+ \sqrt{2 \epsilon ( 1 + \epsilon)}
                       ~^c\! R_{\rm TL}^{00} \cos \phi
                       + \epsilon~^c\!R_{\rm TT}^{00} \cos 2 \phi\right\},\\
\sigma_0P_{x'}^h & = & \frac{|{\bf q}|}{k_w}\left\{
                       \sqrt{2\epsilon(1-\epsilon)}\,^c\!R_{\rm TL'}^{x'0}\cos\phi
                     + \sqrt{1-\epsilon^2} R_{\rm TT'}^{x'0}\right\},\\
\sigma_0P_{y'}   & = & \frac{|{\bf q}|}{k_w}\left\{
                       R_{\rm T}^{y'0} + \epsilon R_{\rm L}^{y'0} \right.\\&&\left.
                     + \sqrt{2\epsilon(1+\epsilon)} ~^c\!R_{\rm TL}^{y'0} \cos\phi
                     + \epsilon ~^c\!R_{\rm TT}^{y'0} \cos2\phi\right\},\\
\sigma_0P_{z'}^h & = & \frac{|{\bf q}|}{k_w}\left\{
                       \sqrt{2\epsilon(1-\epsilon)} \,^c\!R_{\rm TL'}^{z'0}\cos\phi
                     + \sqrt{1-\epsilon^2} R_{\rm TT'}^{z'0}\right\}.
\end{eqnarray*}
In total, a number of 72 polarization observables can be measured in
pseudoscalar meson electroproduction. Since only 36 are different,
each observable can be obtained in two different ways. For the
target-polarization asymmetry the relation $R_{\rm T}^{0y} =
-^c\!R_{\rm TT}^{y'0}$ holds, allowing us to determine the
target-polarization asymmetry by measuring the helicity-independent
recoil polarization $P_{y'}$.

To illustrate the sensitivity of these observables to the leading
multipoles, one drops the small contributions.  For instance, all
longitudinal multipoles and also interferences with them are expected
to be small. Hence, the helicity-independent polarization is dominated
by $R_{\rm T}^{y'0}$ and $^c\!R_{\rm TT}^{y'0}$, which can be written
as
\begin{eqnarray*}
R_{\rm T}^{y'0}        & \approx & \sin\theta\,\Im \left\{E_{0+}^*
                       (3 \cos\theta (E_{2-}-3M_{2-})-2M_{1-})\right\},\\
^c\!R_{\rm TT}^{y'0} & \approx & 3 \sin\theta\cos\theta \,
                       \Im \left\{E_{0+}^*\left(E_{2-}+M_{2-}\right)\right\}.
\end{eqnarray*}
Thus, the interference with $E_{0+}$ amplifies the sensitivity to the
$d$-wave multipoles $E_{2-}$ and $M_{2-}$. In particular, $^cR_{\rm
TT}^{y'0}$ is proportional to the sine of the phase difference $\phi_0
- \phi_2$ between $E_{0+}$ and $E_{2-} + M_{2-}$.  The $\theta$
dependence shows, that the maximum sensitivity to $^c\!R_{\rm
TT}^{y'0}$ is at $\theta=135^\circ$. In this experiment
$\theta=120^\circ$ was chosen as compromise between the sensitivity
and the acceptance of the setup.

The helicity-dependent polarizations are dominated by $|E_{0+}|^2$
\begin{eqnarray*}
R_{\rm TT'}^{x'0} & \approx &
- \sin\theta \left[|E_{0+}|^2-\Re\{E_{0+}^*(E_{2-}-3M_{2-}\}\right],\\
R_{\rm TT'}^{z'0} & \approx & \cos\theta |E_{0+}|^2 -2\Re\{E_{0+}^*[M_{1-}
\\&&-\cos\theta (E_{2-} - 3M_{2-})]\},
\end{eqnarray*}
they also show sensitivity to the longitudinal $S_{0+}$ multipole via
interference with $E_{0+}$
\begin{eqnarray*}
^c\!R_{\rm TL'}^{x'0} & \approx & \cos\theta\,\Re\{S_{0+}^*E_{0+}\},\\
^c\!R_{\rm TL'}^{z'0} & \approx & \sin\theta\,\Re\{S_{0+}^*E_{0+}\}.
\end{eqnarray*}


The experiment was performed at the three spectrometer setup of the A1
collaboration \cite{Blo98} at MAMI-C. The incident electron beam with
an energy of 1508\,MeV and an average current of $10\,\mathrm{\mu A}$
was delivered on a liquid hydrogen target with a length of 5\,cm,
giving a luminosity of $L = 13.4\,\mathrm{MHz/\mu barn}$. The average
polarization of the beam was 79\%.

For the detection of the electron, spectrometer B with a solid angle
acceptance of 5.6\,msr and a momentum acceptance of 15\% was used. The
recoil proton was detected by spectrometer A with a solid angle
acceptance of 21\,msr and a momentum acceptance of 20\%. The electron
arm was set at a central angle of $\theta_{\rm e} =18^\circ$ and a
central scattered electron energy of $E_{\rm e} =
678.4\,\mathrm{MeV}$, defining a photon virtuality of $Q^2 =
0.1\,\mathrm{GeV^2}/c^2$ and a photon polarization of $\epsilon =
0.718$. The proton arm was set at $\theta_{\rm p} = 26.2^\circ$ with a
central momentum of $p_{\rm p} = 660\,\mathrm{MeV}/c$ to detect
protons with an $\eta$ c.m. angle of $\theta_\eta = 120^\circ$ and
$\phi_\eta=0^\circ$.

Spectrometer B was equipped with four layers of vertical drift
chambers for position and angular resolution and two layers of plastic
scintillators for timing resolution and trigger. A gas Cherenkov
detector separated electrons and charged pions. Spectrometer A was
equipped with the same focal plane detectors as B, only the Cherenkov
detector was replaced by a focal plane polarimeter consisting of a
layer of carbon with a thickness of 7\,cm and four layers of
horizontal drift chambers to detect the secondary scattering process
of the recoil protons in the carbon \cite{Pospischil:2000pu}.

The electrons were identified by the Cherenkov detector in
spectrometer B, the protons were identified by the time-of-flight
method. After correction of the coincidence time for the path lengths,
an overall time-of-flight resolution of 1.1\,ns FWHM was achieved. The
$\eta$ production process was identified by the missing-mass spectrum
of the four momentum balance, the $\eta$ peak showed a width of
$1.6\,\mathrm{MeV}/c^2$ FWHM.

After the cut in coincidence time and missing mass the events with a
clearly identified secondary scattering vertex in the carbon analyzer
and a scattering angle of more than $8^\circ$ were selected for the
determination of the recoil polarization. For these events the
azimuthal angle in the polarimeter detector plane $\phi_{\rm FPP}$ was
determined. A simultaneous maximum likelihood fit of the center of
mass polarizations $P_{x'}^h$, $P_{y'}$, and $P_{z'}^h$ was
performed. The statistical errors of the polarizations were determined
from the covariance matrix of the maximum likelihood fit, off-diagonal
elements, \textit{i.e.} correlations, could be neglected.

After the fitting procedure, two correction factors were
applied. First, the acceptance of the spectrometers is large over the
center of mass angular range. The related correction was determined by
using the model Eta-MAID \cite{Chiang:2001as} as input for an event
generator to extract the average polarization of the analysis chain as
compared to the input polarization at nominal kinematics. In this
step, also the radiative corrections were included in the
simulation. The resulting correction was negligible for $P_{z'}^h$ and
$P_{y'}$, while for $P_{x'}^h$ this correction is $\approx 4\%$. The
spread of the polarization components within the acceptance is mainly
caused by the well known angular structure of the cross section, thus
only a minor systematic error is induced by this procedure, which was
estimated by comparison to the same procedure with the use of a simple
phase space isotropic generator to be less than 0.5\% relative.

A second correction factor was applied to account for the background
contribution. After all particle and reaction identification cuts a
background contribution of $\approx 2\%$ by accidental coincidences
(determined by a cut on the side bands of the coincidence time peak)
and a contribution of $\approx 8\%$ by true two pion events with a
missing mass of the two pion system in the region of the $\eta$ mass
remains. The latter can be polarized, so the background polarization
was determined by using the missing-mass region below and above the
$\eta$ missing-mass peak. Both regions showed within the error bars
the same polarization of $\approx -31\%$ for the helicity-dependent
polarizations, so we assume that the background in the region of the
$\eta$ missing-mass peak has the same polarization. The overall
background correction factor is 1.138 for $P_{y'}$ and 1.094 for
$P_{x'}^h$. For $P_{z'}^h$ the correction is small, since the
background polarization is of the same order as the true
polarization. A conservative estimate of 10\% error in the background
polarization leads to a relative systematic error of 1\% in the
background correction factor.

The overall systematic error for the helicity-dependent polarizations
is dominated by the error of the beam polarization, for the
helicity-independent by the uncertainty in analyzing power, spin
precession, and polarimeter efficiency \cite{Pospischil:2000pu}.


The extracted values of the polarization observables are\\
\centerline{
  \begin{tabular}{rcr@{$\,\pm\,$}l@{\,(stat.)$\,\pm\,$}l}
    $P_{x'}^h$ & = & $-67.6$ & 3.2 & 2.6\,(sys.)\%, \\
    $P_{y'}$   & = & $ 16.1$ & 3.2 & 2.3\,(sys.)\%, \\
    $P_{z'}^h$ & = & $-29.3$ & 2.6 & 2.6\,(sys.)\%. \\[1mm]
  \end{tabular}
}

\begin{figure}
\includegraphics[width=\columnwidth]{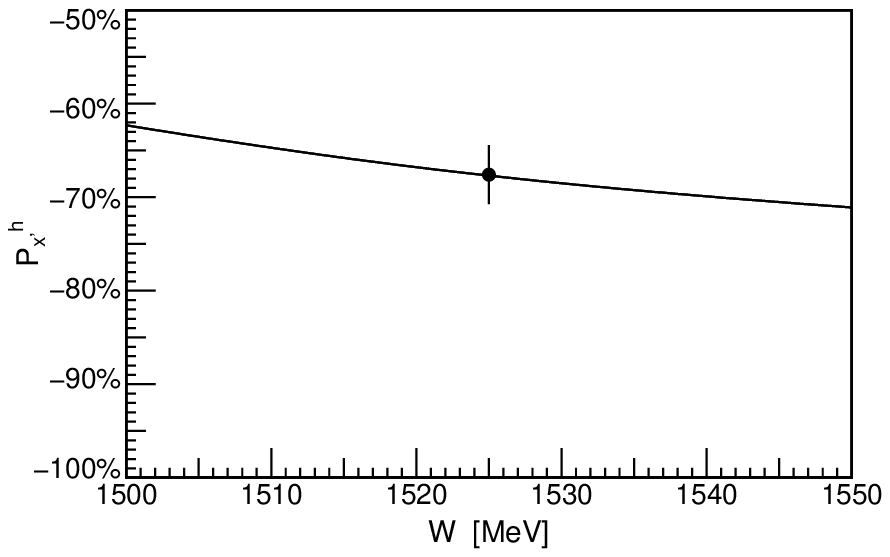}
\includegraphics[width=\columnwidth]{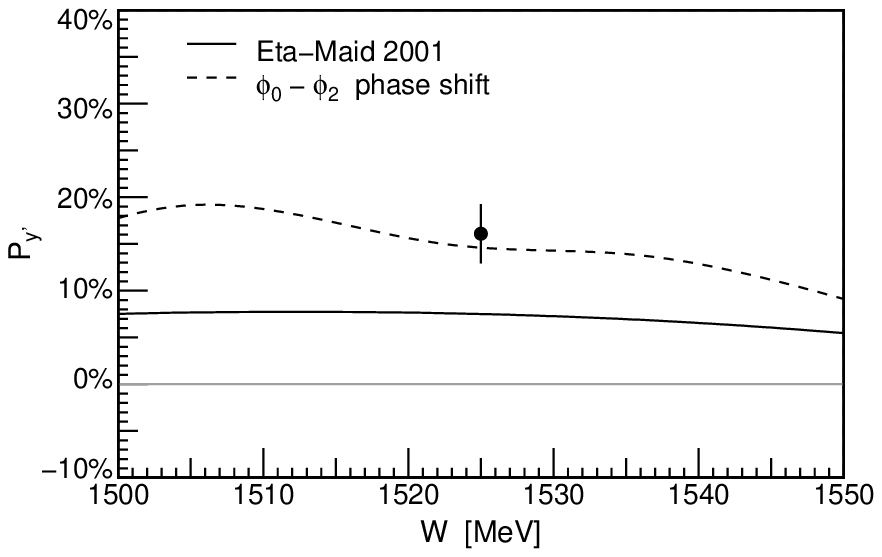}
\includegraphics[width=\columnwidth]{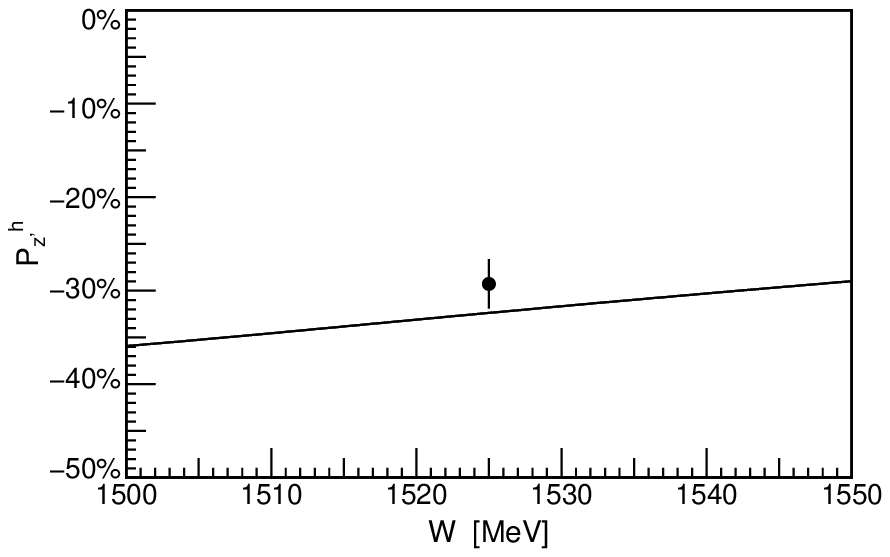}
\caption{ 
  Recoil polarization observables as functions of the c.m.  energy $W$
  at $\theta=120^\circ$, $Q^2=0.1\,\mathrm{GeV}^2/c^2$, and
  $\epsilon=0.718$ (statistical errors only). The solid line shows the
  prediction of Eta-MAID \protect\cite{Chiang:2001as}, the dashed line
  the same model with the energy dependent phase shift of 
  ref.\ \cite{Tiator:1999gr}. The range in c.m. energy $W$ corresponds to
  the acceptance of the experiment, the data were corrected to the
  central point of $W=1525\,\mathrm{MeV}$ as described in the text.}
\label{fig:polarization}
\end{figure}

Fig.\ \ref{fig:polarization} shows the result in comparison with
Eta-MAID over the accepted energy range.

Clearly, the double polarization observables $P_{x'}^h$ and $P_{z'}^h$
are well described by the model. First, this confirms the dominance of
the $s$-wave multipoles in this region, as suggested by the
unpolarized experiments. These observables are, however, also
sensitive to the longitudinal $s$-wave multipole $S_{0+}$, which is
set to zero in the model, since existing $\eta$ production data are
not sensitive enough to justify a finite value.

A first estimate of the longitudinal excitation of the $S_{11}(1535)$
resonance was extracted from pion production \cite{Aznauryan:2004jd}.
A value of 20\% of the transverse amplitude was given in this
reference, an isolated variation of $S_{0+}$ to this value would
change the Eta-MAID prediction \textit{e.g.\ } for $P_{z'}^h$ by 9\%.

The single polarization observable $P_{y'}$ clearly disagrees with the
model (solid line). However, if we apply the strong phase change
between $E_{0+}$ and $E_{2-}+M_{2-}$ discussed in
ref. \cite{Tiator:1999gr}, which was introduced to describe the Bonn
polarized target data \cite{Bock:1998rk}, the data point is in good
agreement with the model. In other words, this data set is consistent
with the Bonn polarized target data, which were excluded from the
standard Eta-MAID fit. Such a strong phase change is not easy to
achieve if one assumes a standard Breit-Wigner behavior for the
$S_{11}(1535)$ resonance. The unitary approaches \cite{Kaiser:1996js,
Inoue:2001ip} could show in principle this phase variation by coupled
channel effects. Ref.\ \cite{Kaiser:1996js} predicts however a
strength of $S_{0+}$ of nearly 30\% of the $E_{0+}$ strength, which
manifestly contradicts our double-polarized results.

Clearly, a broader basis of polarization data with large angular
coverage is needed to further clarify the nature of the
$S_{11}(1535)$ resonance. Further experiments on polarization
observables are planned by different groups in the near future.

\acknowledgments{
  The authors like to thank the MAMI accelerator group for their
  extraordinary commitment to this first experiment with MAMI-C. This
  work was supported by the Federal State of Rhineland-Palatinate and
  by the Deutsche Forschungsgemeinschaft with the Collaborative
  Research Center 443.
}

\end{document}